\documentclass{article}
\usepackage{graphicx}

\begin{document}                                 
\noindent {\large Lamb Mossbauer factor using non-extensive Statistics}
\vskip1.5cm
\noindent {\bf Ashok Razdan}

\noindent {\bf Nuclear Research Laboratory }

\noindent {\bf Bhaba Atomic Research Centre }

\noindent {\bf Trombay, Mumbai- 400085 }
\vskip 1.5cm
\noindent {\bf Abstract :}

Lamb Mossbauer Factor is derived using Tsallis Statistics. Recoil free factor so
obtained, has weak temperature dependence which is similar to the
temperature dependence observed  for Lamb Mossbauer factor 
of a fractal.

\noindent {\bf Introduction:}

Anomalous behavior observed in properties of various physical 
system have been explained by using non-extensive   Tsallis statistics. These 
physical systems /phenomena include turbulence in plasma [1], Cosmic ray background 
radiation [2], self gravitating systems  [3], econo-physics[4], electron - 
positron annihilation [5],classical and quantum chaos [6], linear response 
theory [7], Levy type anomalous super diffusion [8], thermalization of 
electron - phonon systems [9], low dimensional dissipative systems [10], etc.

  It has been shown that non-extensive features get manifested in those systems 
which have long range forces, long memory effects or in those systems which 
evolve in (non Euclidean like space-time) fractal space time [11 and reference therein]. Non - 
extensive statistics is based on two postulates [11 and reference therein]. First is the definition of non-
extensive entropy
\begin{equation}
S_q  = \frac{ 1- P_i^{q}}{q -1}                              
\end{equation}

where q is 
the non-extensive entropic index, k is a positive Constant and $P_i$ are the 
probabilities of the microscopic states with $\sum P_i$  = 1.

The second postulate is the definition of  energy $U_q$  = $\sum P_i^{q} E_i , $ 
where $E_i$ is the energy spectrum. 
As q $\rightarrow$1, $S_q$  = - $\sum p_i ln P_i $, which is the Boltzman Gibbs Shanon 
entropy.

Apart from other applications it has  been suggested [11] that non-extensive
statistics can be applied to complex systems like glassy materials and fractal /
multi-fractal or unconventional structures also.
Anomalous Lamb Mossbauer factor behavior observed in glasses and biological 
system have motivated us to use non-extensive statistics. We will also compare 
Lamb Mossbauer factor using Tsallis statistics with Lamb Mossbauer factor of
a fractal.

\noindent  {\bf  Anomalous f - factor:}

Lamb Mossbauer factor or recoil free factor f is given by
\begin{equation}
log f = \frac{ -4 \pi^{2} < x_m ^{2} >}{\lambda^{2}}
\end{equation} 

Where $x_m ^2$ is mean square displacement of Mossbauer atom of mass m, 
averaged over  the nuclear life time. For Debye model of lattice vibrations, the 
Lamb Mossbauer factor can be written as [12]  

\begin{equation}
f=exp[- \frac{-6 R }{k {\theta_D}} (\frac{1}{4} + (\frac{T}{\theta_D})^2 \int_{0}^{\frac{\theta_D}{T}} \frac {x dx}{exp(x)-1})]  
\end{equation}

Where $x$ = $\frac{\hbar \omega}{kT}$, R is recoil energy and $\theta_D$ is the
Debye temperature defined by $\hbar \omega_{max}$= k $\theta_D$.

Equation (3) has been obtained using Boltzman Gibbs statistical mechanics.
f-factor is a very important parameter to study temperature dependence of 
dynamics of Mossbauer atom. In most of the cases equation (3) explains 
the f-factor experimental data very effectively. However, there are various system 
like superconductors, glasses, biological system etc. where f - factor 
experimental data can not accounted by equation (3). Anomalous 
behavior of experimental f - factor data in various types of superconductors have 
been explained by incorporating anharmonicity in the dynamics of 
Mossbauer atom [13,14]. For many biological systems e.g. in deoxy -myoglogin 
above 265 K, f - factor experimental data [15,16]can not explained by equation 
(3). Again in various types of glassy materials f-factor experimental data [17] shows 
temperature dependence similar to biological systems. 

\noindent {\bf Theory :}

To derive non-extensive form of Lamb Mossbauer factor we have to use 
non-extensive quantum distribution function of bosons. It is very 
difficult to derive exact analytical expression for non-extensive distribution 
function. However, there are many studies which provide the approximate 
form of non-extensive distribution functions [2,18]. In the present paper we use 
dilute gas approximation (DGA) for boson distribution function. For DGA 
case, the average occupation number is given as [18]
\begin{equation}
< n_q > = \frac{1}{(1+(q-1)\beta (E_i -\mu)^{\frac{1}{q-1}}-1)}
\end{equation}

When dealing with system in contact with the heat bath at the temperature 
$\beta$, we have 
\begin{equation}
< n_q > =\frac{\hbar \omega_i}{(1+(q-1) \frac{\hbar\omega_i}{kT})^{\frac{1}{q-1}}-1}
\end{equation}
 
If $x_m^2$ is the mean square displacement of Mossbauer atom of mass m, 
and for  the case of harmonic motion, we have
\begin{equation}
3N m {x_m} ^{2} \omega_i ^{2} = ( n_q + \frac{1}{2})\hbar \omega_i 
\end{equation}
The harmonic approximation in (6) means that each of 3N as oscillators 
modes have $\delta$ function distribution. Let $f(\omega_i)$ be the spectral distribution 
of the lattice Vibration frequencies of the lattice. Let us find average of  
$4 \pi^2 x_m ^2$ over the entire vibration spectrum.
\begin{equation}
\frac{-4 \pi^{2} < x_m ^2 >}{\lambda^{2}}= \int_{0}^{\omega_{max}} \frac{\hbar \omega_i}{m \omega_i ^2}( \frac{1}{2}+n_q) \frac{4\pi^2 f(\omega_i) d \omega_i}{3N \lambda^2}     
\end{equation}

For the case of Debye approximation    
\begin{equation}
f(\omega_i)= 9N \frac{\omega_i ^2}{\theta_D ^2}  \frac {\hbar^3}{k^3}
\end{equation}
where $\hbar w_{max}$= k $\theta_D$.

Using equations (3) and (7) and (8), we have  

\begin{equation}
f= exp \left\{- \frac{E^2_\gamma}{3mc^2} \int^{\omega_{max}}
_0 \left(\frac{1}{2} + n_q \right) 9
\left\{\frac{\hbar}{k}\right\}^3 \frac{\omega^2_i}{\theta^3_D}\frac{d\omega_i}{\hbar \omega_i}\right\}                                                                                                  
\end{equation}
                                                                                         
Equation (9) represents Lamb Mossbauer factor using Tsallis statistics 
where $n_q$  is given by equation (5)

\begin{figure}
\begin{center}
\includegraphics[angle=270,width=7.0cm]{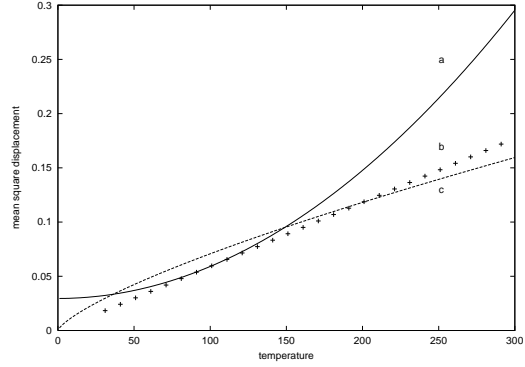}
\caption{Lamb Mossbauer factor v/s temperature (a) equation 3, (b) equation 9 and (c)equation 10.} 
\end{center}
\end{figure}

\noindent {\bf Results and Discussion :}

We have numerically solved equation  (9) for q = 2 shown as ( + ) corresponding to
curve 'b' of figure 1. The curve 'c' of figure 1  
corresponds to Lamb Mossbauer factor of a fractal. It has been shown 
that f factor for a fractal is given by [19,20]     

\begin{equation}
log f \propto T^{(0.74+\frac{4}{3})}  exp ( - c  x^{0.74} T^{0.74})
\end{equation}

We have plotted  proportional temperature dependence of equation (10)
(Lamb Mossbaur factor of a fractal) in curve 'c'. 
curve 'a' in this figure has been obtained using equation (3), 
corresponding  to Lamb Mossbaur factor using Boltzman Gibbs statistical 
mechanics. It is clear from figure 1 that Curve 'b' and 'c' have similar or 
overlapping behavior. Comparison of Curve 'a' on one hand, to Curves 'b' 
and 'c' on the other hand makes it clear that the Lamb Mossbauer fraction 
in non-Euclidean case is smaller than in the case of Euclidean solid.
It is also clear that both 
fractal and non- Euclidean Lamb Mossbauer fraction  is less temperature 
dependent than fraction for  Euclidean solids.
The similarity between Curve 'b' and 'c' is expected  because 
non-extensive statistics corresponds to non-Euclidean space time which 
is fractal in nature.

\begin{figure}
\begin{center}
\includegraphics[angle=270,width=7.0cm]{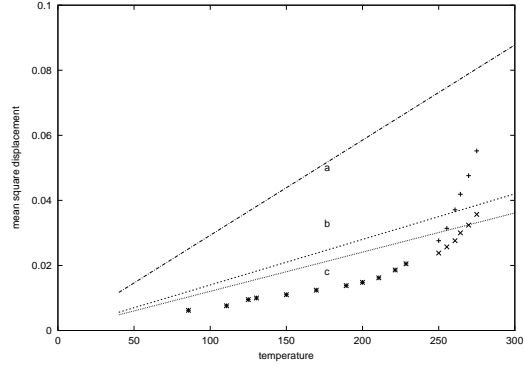}
\caption{Lamb Mossbauer factor defined in equation 10 for (a) q=3.0, (b)q =1.2 and (c) q=0.6.
Experimental data has been taken from reference 17.} 
\end{center}
\end{figure}

	In Fig. 2 we are plotting experimental data corresponding to Fec$l_2$ 
dissolved in glycerol-water mixture in super cooled liquid state. The 
experimental data has been taken from  reference [17 ]. The three 
curves corresponding to non-extensive Lamb Mossbauer factor for different 
values of parameter q. It is clear that temperature dependence of glassy 
materials above some characteristic temperature is very strong whereas  non-extensive 
Lamb Mossbauer factor has continuous weak temperature dependence. Thus non-
extensive f-factor cannot account for anomalous f - factor data of glasses 
and bio systems. It is interesting to mention here for the sake of completeness 
that Lamb Mossbauer factor using generalized algebra  (of deformation 
concepts ) has been derived earlier and deformation parameter q has 
been identified as anharmonic coefficient when applied to experimental 
Mossbauer data [13,14]. Now with the advent of generalized statistics, Lamb 
Mossbauer factor seems to behave more like an f-factor of a fractal. It will 
be interesting exercise to use both generalized algebra and generalized 
statistics to derive Lamb Mossbauer factor. Combination of generalized statistics
and generalized algebra has been successfully applied to the study of COBE [21]results. 

\noindent {\bf Conclusion :}

In this paper we have derived Lamb Mossbauer factor for 
generalized statistics and found it very similar to f-factor of a fractal.
	
\noindent {\bf References: }
\begin{enumerate}
\item B.M.Boghosian, Phys. Rev. E 53(1996)4745
\item C.Tsallis, F.C. Sa Barreto, E.D.Loh, Phys. Rev. E 52(1995) 1447
\item V.H.Hamity, D.E.Barraco, Phys. Rev. Lett. 76(1996)4664
\item C.Tsallis, C.Anteneodo, L.Borland and R.Osorio, Cond-mat/0301307
\item I.Bediaga, E.M.F.Curado and j. Miranda, Physica A 286(2000)156
\item C.Tsallis, A.R.Pastino and W. -M.Zheng, Choas,Solitons and Fractals 8(1997)885,
      Y.Weinstein, S.Lloyd and C.Tsallis, Phys. Rev. Lett. 89(2002)214101
\item A.K.Rajgopal, Phys. Rev. Lett. 76(1996) 3496
\item C.Tsallis, S.V.F. Levy, A.M.C.Souza, R.Maynard, Phys. Rev. Lett. 75(1995)3589
\item I.Koponen, Phys. Rev. E 55(1997)7759
\item M.L.Lyra, C.Tsallis, Phys. Rev. Lett. 80(1998)53
\item C.Tsallis, Physica A 221(1995)277-290  
\item V.G.Bhide in Mossbauer Effect and its Applications, Tata McGraw-Hill Publishing co.Ltd.
      New Delhi 1973 edition
\item Ashok Razdan, Pramana -Journal of Physics, Indian academy of Sciences 54(2000)871
\item Ashok Razdan, Hyperfine Interactions ,122(1999)309
\item I.Nowik, E.R.Bauminger, S.G.Cohen, S.Ofer, Phys. Rev.A 31(1983)2291
\item Ashok Razdan, Eur. Phys. J. B 8(1999)143
\item G.U.Nienhaus,H.Frauenfelder and F.Parak Physical Review B 43(1991)3345
\item Q.A.Wang,A.Le Mehaute, Physics Letters A 242(1998)301
\item K.N.Shrivastava, J.Phys. C: Solid State Phys. 19(1986)L389
\item R.Haung,Y.Hsia and R.Liu, Hyperfine Interactions 55 (1990)1141
\item A.B.Pinheiro,I.Roditi, Physics Letters A 242(1998)296
\end{enumerate}
\end{document}